**Anomalous Hall Effect and Perpendicular Magnetic Anisotropy in Ultrathin Ferrimagnetic NiCo$_2$O$_4$ Films**

Xuegang Chen,[1,2*] Qiuchen Wu,[1*] Le Zhang,[1] Yifei Hao,[1] Myung-Geun Han,[3] Yimei Zhu,[3] Xia Hong[1,a)]

[1] Department of Physics and Astronomy & Nebraska Center for Materials and Nanoscience, University of Nebraska–Lincoln, Lincoln, NE 68588-0299, USA

[2] Institutes of Physical Science and Information Technology, Anhui University, Hefei 230601, People's Republic of China

[3] Condensed Matter Physics and Materials Science Department, Brookhaven National Laboratory, Upton, NY 11973-5000, USA

[*] These authors contributed equally

[a)] Author to whom correspondence should be addressed: xia.hong@unl.edu

**Abstract**

The inverse spinel ferrimagnetic NiCo$_2$O$_4$ possesses high magnetic Curie temperature $T_C$, high spin polarization, and strain-tunable magnetic anisotropy. Understanding the thickness scaling limit of these intriguing magnetic properties in NiCo$_2$O$_4$ thin films is critical for their implementation in nanoscale spintronic applications. In this work, we report the unconventional magnetotransport properties of epitaxial (001) NiCo$_2$O$_4$ films on MgAl$_2$O$_4$ substrates in the ultrathin limit. Anomalous Hall effect measurements reveal strong perpendicular magnetic anisotropy for films down to 1.5 unit cell (1.2 nm), while $T_C$ for 3 unit cell and thicker films remains above 300 K. The sign change in the anomalous Hall conductivity ($\sigma_{xy}$) and its scaling relation with the longitudinal conductivity ($\sigma_{xx}$) can be attributed to the competing effects between impurity scattering and band intrinsic Berry curvature, with the latter vanishing upon the thickness driven metal-insulator transition. Our study reveals the critical role of film thickness in tuning the relative strength of charge correlation, Berry phase effect, spin orbit interaction, and impurity scattering, providing important material information for designing scalable epitaxial magnetic tunnel junctions and sensing devices using $NiCo_2O_4$.

Magnetic thin films with high spin polarization, high Curie temperature, and perpendicular magnetic anisotropy (PMA) are important material solutions for developing high speed energy-efficient spintronics.[1,2] A promising material candidate is the ferrimagnetic inverse spinel $NiCo_2O_4$ (NCO), where the majority and minority spins are associated with the tetrahedral site Co ions and octahedral site Ni ions, respectively [Fig. 1(a) inset]. Previous studies have shown that epitaxial NCO films exhibit intriguing magnetotransport properties,[3-5] strain tunable magnetic anisotropy,[6] and spin polarization as high as -73%.[7] Being lattice matched with the high performance tunnel barrier candidate, $MgAl_2O_4$, it can be utilized as the spin injection layer for epitaxial magnetic tunnel junctions (MTJ),[7-10] which can potentially suppress the band-folding effect and lead to enhanced tunneling magnetoresistance.[11] Compared with amorphous MTJs, epitaxial $NCO/MgAl_2O_4/NCO$ heterostructures also facilitate the understanding of symmetry-based spin filtering effect.[12] Despite the emerging interests in epitaxial NCO thin films,[2-7, 13-18] fundamental understanding of the thickness scaling behavior of its intriguing magnetic properties remains elusive.

In this work, we report the observation of unconventional anomalous Hall effect (AHE) and strong PMA in ultrathin (001) NCO films strained on MgAl$_2$O$_4$ substrates. The Curie temperature ($T_C$) of the NCO films remains above 300 K in films as thin as 3 unit cell (uc), while PMA is sustained even in 1.5 uc films. The anomalous Hall (AH) resistance shows a temperature-driven sign change, evolving from clockwise to counterclockwise hysteresis with decreasing temperature. The scaling behavior between the AH conductivity ($\sigma_{xy}$) and longitudinal conductivity ($\sigma_{xx}$) points to the collective contributions of spin-dependent scattering and band intrinsic Berry phase effect to the AHE in the metallic phase, with the latter vanishing upon the thickness driven metal-insulator transition (MIT).

We deposited epitaxial NCO films on (001) MgAl$_2$O$_4$ substrates via off-axis radio frequency magnetron sputtering at 320 °C in 100 mTorr processing gas (Ar:O$_2$ = 1:1).[3] X-ray diffraction (XRD) measurements reveal (001)-oriented growth with no impurity phase [Fig. 1(a)]. The Laue oscillations around the Bragg peaks were used to calibrate the growth rate. The *c*-axis lattice constant is 8.20 Å, higher than the bulk value (8.13 Å), which is consistent with a strained film on MgAl$_2$O$_4$ ($a$ = 8.08 Å).[13, 16] The high crystallinity of the samples is clearly revealed by the high-angle annular dark-field scanning transmission electron microscopy (HAADF-STEM) and selected area electron diffraction (SAED) [Fig. 1(b)]. The as-grown NCO films show smooth surface morphology with a typical surface roughness of 2-3 Å [Fig. 1(c)]. Magnetotransport studies were carried out in van der Pauw device geometry using a Quantum Design Physical Property Measurement System connected with Keithley 2400 SourceMeter or standard lock-in technique (SR830) with low excitation current (≤10 µA). We studied two samples for each thickness and the results are highly consistent.

Figure 1(d) shows the temperature dependence of the sheet resistance $R_\square = \rho_{xx}/t_{\text{NCO}}$, where $\rho_{xx}$ is the longitudinal resistivity and $t_{\text{NCO}}$ is the film thickness, of 1.5-5 uc NCO films. For the 3-5 uc samples, $R_\square$ shows a moderate, metallic $T$-dependence ($dR/dT > 0$) below 350 K, consistent with that of thick NCO films.[3, 13, 14] A resistance upturn appears ($dR/dT < 0$) at low temperature, with the upturn temperature ($T_{\text{up}}$) increasing with reducing film thickness. This moderate resistance upturn is the signature behavior of weakly localized conductors, which can be due to electron back-scattering induced quantum interference effect or enhanced electron-electron interaction.[19] The 1.5 uc film becomes insulating over the entire temperature range, with $R_\square(T)$ well described by variable range hoping. The transition to the strongly localized behavior occurs in the 2 uc film as $R_\square$ approaches $\frac{h}{e^2}$ ~25.9 kΩ. This thickness-driven MIT has been widely observed in correlated oxide thin films, which may originate from dimensionality crossover,[19, 20] enhanced correlation induced energy gap in ultrathin films,[19, 21] or variation in oxygen octahedral distortion[22] and impurity/defect states[17] at surfaces/interfaces.

Figure 2(a) shows the AHE hysteresis taken in 1.5-5 uc NCO films at 100 K. The Hall resistivity, $\rho_{xy} = R_{xy}t_{\text{NCO}}$ with $R_{xy}$ the Hall resistance, can be decomposed into the normal Hall and AHE contributions: $\rho_{xy} = R_0 H + \rho_{\text{AHE}}$. Here the normal Hall coefficient $R_0 = 1/en_{\text{eff}}$ yields the effective carrier density $n_{\text{eff}}$, $H$ is the out-of-plane magnetic field, and $\rho_{\text{AHE}}$ is the AH resistivity. All samples exhibit counterclockwise AHE hysteresis, which is defined as positive. The coercive field ($H_c$) is significantly reduced in the 1.5 and 2 uc films [Fig. 2(b)], which may be attributed to domain nucleation and domain wall motion promoted by interfacial defects[23] and/or enhanced thermal fluctuation with reduced dimension. The sharp switching of the AHE hysteresis suggests that PMA is preserved in NCO films as thin as 1.5 uc. This can be well accounted for by the 0.6% compressive strain for (001) NCO on MgAl$_2$O$_4$, which yields ~1 meV magnetic anisotropic energy between the in-plane and out-of-plane orientations, well exceeding the shape anisotropy (~μeV).[3, 6, 14] From the temperature dependence of $\rho_{AHE}$, we found that $T_C$ is above 300 K for the 3-5 uc NCO films and decreases to 230 K for the 2 uc (1.6 nm) NCO and 170 K for the 1.5 uc (1.2 nm) film [Fig. 2(c)]. Figure 2(d) compares the $T_C$ at the ultrathin limit for epitaxial thin films of magnetic complex oxides, such as $SrRuO_3$ (SRO)[24] and $La_{0.7}Sr_{0.3}MnO_3$ (LSMO),[25] and two-dimensional (2D) van der Waals (vdW) magnets, including $VSe_2$,[26] $MnSe_2$,[27] $CrSe_2$,[28] $CrI_3$,[29] $CrCl_3$,[30] $Fe_3GeTe_2$ (FGT),[31, 32] and $Cr_2Ge_2Te_6$ (CGT).[33] Both classes of material systems are well suited to build crystalline heterostructures based on structurally similar yet functionally distinct constituents.[34, 35] The combined high $T_C$ and strong PMA of ultrathin NCO films outperform most magnetic complex oxides. Compared to 2D vdW magnets, it offers not only highly competitive material parameters but also the distinct advantage of scalable growth, where high quality epitaxial ultrathin films can be controlled with atomic precision.[36] These features make NCO a highly promising material candidate for developing nanoscale multifunctional applications.

One intriguing property of NCO is the AHE exhibits a nonmonotonic $T$-dependence that leads to temperature/film thickness driven sign change in $\rho_{AHE}$.[3] Figure 3(a) shows $R_{xy}$ of a 2 uc film at various temperatures. Below $T_C$ of 230 K, $R_{xy}$ exhibits a non-linear $H$-dependence, signaling the appearance of AHE. The magnitude of AH resistance $R_{AHE} = \rho_{AHE}/t_{NCO}$ first increases below $T_C$, which can be attributed to the initial rapid growth of magnetization. After reaching a local maximum at 190 K, $R_{AHE}$ starts to decrease with decreasing temperature, and crosses zero at $T_{sc} = 150$ K. Similar $T$-dependence of $R_{AHE}$ has been observed in all ultrathin films [Fig. 3(b)]. The transition from the negative (clockwise) to positive (counterclockwise) hysteresis occurs smoothly without any abrupt jump in $R_{AHE}$ [Fig. 3(c)], and the sign change temperature $T_{sc}$ does not exhibit an apparent dependence on film thickness [Fig. 2(c)]. As NCO is a multiband conductor,[15] we also examined $n_{\text{eff}}$ from the high field (up to 50 kOe) normal Hall effect. Figure 3(d) reveals a transition from electron-like ($R_0 < 0$) to hole-like ($R_0 > 0$) behavior upon cooling for the 2 uc sample. The transition temperature is about 200 K, much higher than $T_{\text{sc}}$, ruling out the change of carrier type as the origin of $R_{\text{AHE}}$ sign change. We also extracted $H_c$ from the AHE hysteresis. The exponential $T$-dependence [Fig. 3(e)] points to thermally activated domain wall depinning.[23] The smooth $T$-dependences of $R_{\text{AHE}}$ and $H_c$ [Fig. 3(f)] indicate that the sign change in AHE is not induced by an abrupt change of magnetic state in the samples.

Previous studies of thick NCO films show that the sign change phenomenon can result from the competition between multiple AHE mechanisms.[3] Based on quantum transport theory of multiband magnetic conductors, the scaling relation between $\sigma_{xy}$ and $\sigma_{xx}$ depends on the relative strength of impurity scattering rate with respect to the Fermi energy:[37]

$$\sigma_{xy} = \text{constant} ; \tag{1a}$$

$$\sigma_{xy} \propto \sigma_{xx}^{1.6}. \tag{1b}$$

Equation (1a) describes the scattering-independent band intrinsic Berry phase contribution to AHE, which dominates in moderately dirty metals.[37] Equation (1b) depicts the impurity scattering dominated behavior in dirty metals.[37, 38] For thick NCO films, $\sigma_{xx}$ is close to the boundary between these two regions ($\sigma_{xx} \sim 10^3\ \Omega^{-1}\text{cm}^{-1}$), and both mechanisms can coexist. For super-clean metals ($\sigma_{xx} \gtrsim 10^6\ \Omega^{-1}\text{cm}^{-1}$), the skew scattering mechanism arises due to the spin-orbit interaction associated with impurities, which yields a linear dependence between $\sigma_{xx}$ and $\sigma_{xy}$.[37] This mechanism is less likely to be prominent in NCO, which is prone to formation of various disorders due to its multi-cation, multivalence nature.[2, 4, 15-18]

Figure 4(a) shows $\sigma_{xy}$ vs. $\sigma_{xx}$ for the ultrathin NCO films, where $\sigma_{xx}$ is extracted from the zero-field $R_\square(T)$ [Fig. 1(d)] and $\sigma_{xy} = \rho_{\text{AHE}}/\rho_{xx}^2$. A clear power-law dependence is present in all samples in the intermediate temperature range with the exponent close to 1.6. We fitted the scaling relation by considering both the impurity scattering and band intrinsic contributions:

$$\sigma_{xy} = A_1 \sigma_{xx}^{1.6} + \sigma_{xy}^{(0)}. \tag{2}$$

Here $A_1$ is the fitting parameter and $\sigma_{xy}^{(0)}$ is the Berry phase contribution.[37-40] The band intrinsic nature of the latter has been confirmed in thick films from the scaling behavior between $\sigma_{xy}$ and magnetization.[5] For the 3-5 uc films in the metallic region above $T_{\text{up}}$, the power law scaling between $\sigma_{xy}$ and $\sigma_{xx}$ can be well depicted by Eq. (2). The sign of $\sigma_{xy}$ is thus determined by the relative strength of the positive contribution from dirty metal scattering and the negative Berry phase contribution. With reducing film thickness, $\sigma_{xx}$ decreases, moving further away from the boundary to the moderately dirty region. This is consistent with the rising contribution from impurity scattering ($A_1$), while the intrinsic contribution $\sigma_{xy}^{(0)}$ only shows a slight increase [Fig. 4(b)]. Below $T_{\text{up}}$, $\sigma_{xy}$ exhibits very weak $T$-dependence. In addition to the band intrinsic effect, spin-orbit interaction in impurity scattering, known as the side jump effect $\sigma_{xy}^{(1)}$, may also lead to the scattering insensitive AHE in this region. Overall, the AHE for films of this thickness range (3-5 uc) is qualitatively similar to that observed in thicker films.[3]

The 1.5-2 uc NCO films, on the other hand, exhibit distinct $\sigma_{xy}$ vs. $\sigma_{xx}$ scaling relation. The 2 uc film exhibits a MIT as $R_\square$ exceeds $\frac{h}{e^2}$. While the high temperature AHE scaling still follows Eq. (2), there is a sharp increase in $A_1$. Below $T_{\text{up}}$, $\sigma_{xx}$ falls rapidly and the AHE can be well described by the dirty metal scaling relation alone:

$$\sigma_{xy} = A_2 \sigma_{xx}^{1.6}. \tag{3}$$

Similar scaling is observed in the insulating, 1.5 uc sample, where the negligible contribution of the Berry phase term is corroborated by the facts that $T_{sc}$ approaches $T_C$ [Fig. 2(c)] and the sign change behavior is hard to resolve [Fig. 3(c)]. Equation (3) is qualitatively different from the $\sigma_{xx}$-independent $\sigma_{xy}$ observed in the weakly localized regime of thicker films below $T_{up}$, and can be well accounted for by phonon-assisted hopping and percolative transport.[19, 41, 42] It further shows that the spin scattering source is likely associated with the surface/interface of the sample, whose contribution becomes dominant in ultrathin films.

Based on this scenario, we summarized the AHE mechanisms observed in the ultrathin NCO films as well as previously studied thick films.[3] Figure 4(c) shows the temperature-film thickness diagram for the electronic and magnetic state of 1.5-30 uc NCO films. For 2 uc and thicker films, the intermediate temperature region between $T_C$ and $T_{up}$ corresponds to a metallic phase, and the AHE depends on the collective contributions of band intrinsic effect and impurity scattering. Below $T_{up}$, $R_{AHE}$ becomes independent of scattering, which can be due to the Berry phase and/or side jump effect. Upon the onset of strong localization ($R_\square > \frac{h}{e^2}$),[19, 43] the Berry phase effect vanishes, and the dirty metal scaling behavior dominates $R_{AHE}$. The rich electronic/AHE diagram is thus a clear manifestation of the complex energy landscape in NCO associated with magnetic exchange, electron correlation, spin-orbit interaction, and impurity scattering. The film thickness offers an effective control parameter to tune the relative strength between these energy scales, which can be utilized to design application-specific magnetic properties, including the magnetic transition temperature, coercive field, and the strength and sign of AH effect.

In summary, we show that epitaxial NCO films as thin as 1.5 uc (1.2 nm) can sustain strong PMA, competitive $T_C$ of 170 K, and robust AH effect. The AHE reveals the complex interplay between band intrinsic Berry phase effect and impurity scattering, while the film thickness can be leveraged to tune their relative strength. Our study also paves the path for scalable development of NCO based epitaxial MTJs and sensor devices.

**Acknowledgements**

This work was primarily supported by National Science Foundation through Grant No. DMR-1710461 and EPSCoR RII Track-1: Emergent Quantum Materials and Technologies (EQUATE), Award OIA-2044049. Work by X. Chen was partially supported by the National Natural Science Foundation of China (Grants No. 12104005). Work by M.H. and Y.Z. was supported by the Materials Science and Engineering Divisions, Office of Basic Energy Sciences of the U.S. Department of Energy under Contract No. DESC0012704. TEM sample preparation using FIB was performed at the Center for Functional Nanomaterials, Brookhaven National Laboratory. The research was performed in part in the Nebraska Nanoscale Facility: National Nanotechnology Coordinated Infrastructure and the Nebraska Center for Materials and Nanoscience, which are supported by the National Science Foundation under Award ECCS: 2025298, and the Nebraska Research Initiative.

**Conflict of Interest**

The authors have no conflicts to disclose.

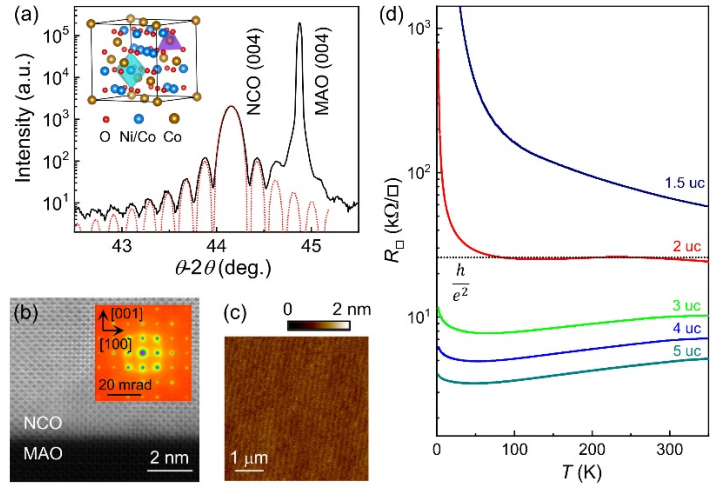

**FIG. 1.** (a) XRD $\theta$-$2\theta$ scan of a 60 uc NCO film on MgAl$_2$O$_4$ (MAO) with a fit to Laue oscillations near the (004) peak (dotted line). Inset: Schematic of NCO crystal structure. (b) HAADF-STEM and (inset) SAED images taken on a 17 uc NCO. (c) AFM topography image of a 5 uc NCO showing atomically flat terraces. (d) $R_\square(T)$ for 1.5-5 uc films. The dotted line illustrates the two-dimensional quantum resistance $\frac{h}{e^2}$.

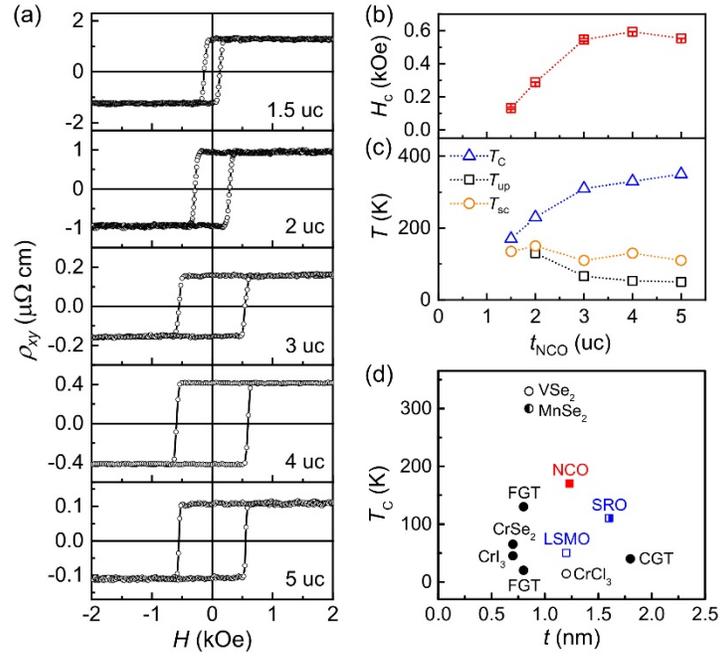

**FIG. 2.** (a) $\rho_{xy}$ vs $H$ hysteresis at 100 K for 1.5-5 uc NCO. (b-c) $t_{NCO}$-dependence of (b) $H_C$ at 100 K, and (c) $T_C$, $T_{up}$, and $T_{sc}$. (d) $T_C$ vs. thickness for epitaxial complex oxide films (squares)[24,25] and 2D vdW magnets (circles).[26-33] Solid symbols: PMA. Open symbols: in-plane magnetic anisotropy.

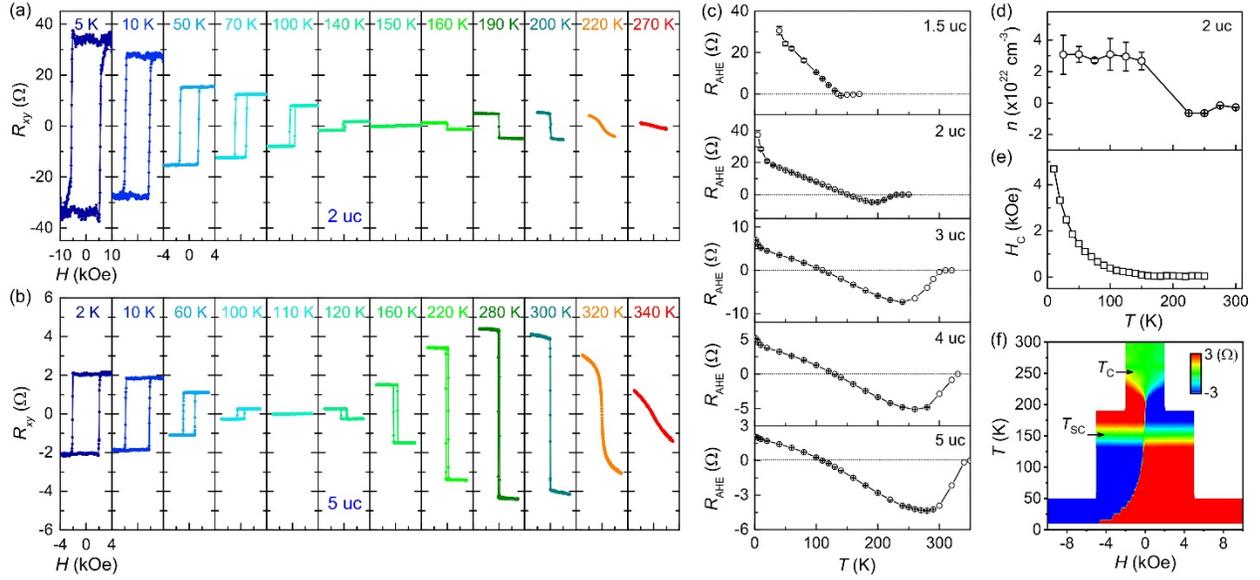

**FIG. 3.** (a) AHE of 2 uc NCO at various temperatures. Field range: ±10 kOe for 5-10 K and ±4 kOe for higher temperatures. (b) AHE of 5 uc NCO at various temperatures. (c) $R_{AHE}$ *vs. T* for 1.5-5 uc NCO. (d) *n vs. T* and (e) $H_c$ *vs. T* for 2 uc NCO. (f) $R_{xy}$ *vs. T* and *H* (sweeping down) for 2 uc NCO.

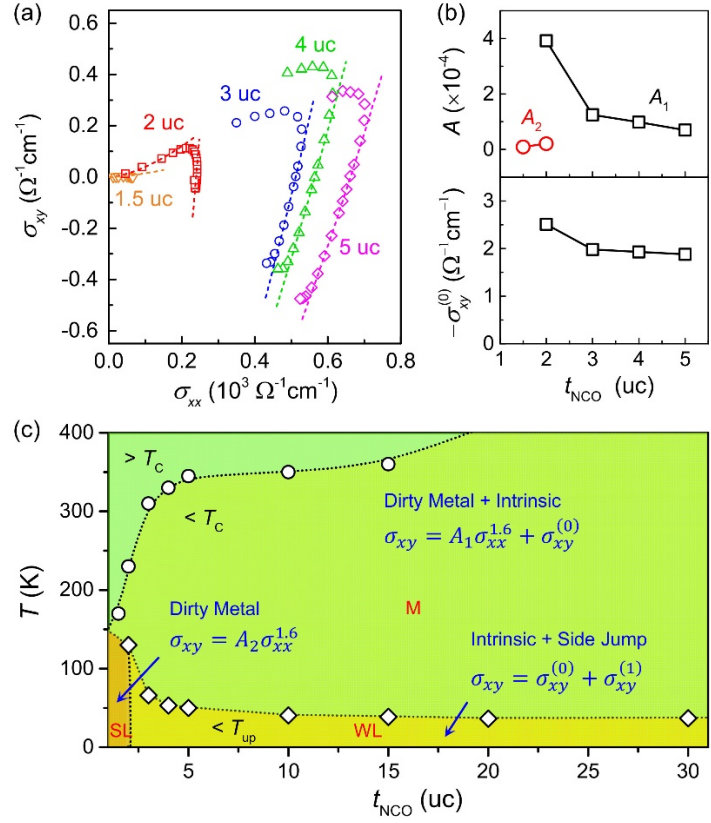

**FIG 4.** (a) $\sigma_{xy}$ vs. $\sigma_{xx}$ for 1.5-5 uc NCO films with fits to Eqs. (2)-(3) (dash lines). (b) $t_{NCO}$-dependence of $A_1$, $A_2$ and $\sigma_{xy}^{(0)}$. (c) Temperature-thickness diagram of the AHE scaling behavior. M: metallic. WL: weakly localized. SL: strongly localized. The data for 10-30 uc films are taken from Ref. [3].